\documentclass[ ]{aa}
\usepackage{psfig}

\usepackage{graphics}
\usepackage{psfig}
\begin{document}
\thesaurus{08(08.14.2;08.09.2;02.12.2)}

\title{First results of UVES at VLT: revisiting RR Tel  
\thanks{Based on 
public data released from the UVES commissioning at the VLT Kueyen
telescope, European Southern Observatory, Paranal, Chile.
}}
\author{
Pier Luigi Selvelli\inst{1}
\and 
Piercarlo Bonifacio\inst{2}
}
\offprints{P.L. Selvelli}
\institute{
C.N.R.--G.N.A.--
Osservatorio Astronomico di Trieste,Via G.B.Tiepolo 11, 
I-34131 Trieste, Italia
\and
Osservatorio Astronomico di Trieste,Via G.B.Tiepolo 11, 
I-34131 Trieste, Italia
}
\mail{selvelli@ts.astro.it}
\date{received .../Accepted...}
\authorrunning{Selvelli \& Bonifacio}
\titlerunning{UVES observations of RR Tel}
\maketitle

\begin{abstract}
We present here the first results of UVES observations of RR Tel.
The  exceptional performances of the instrument have allowed  the
detection  of new spectral features  and have led to an improvement in
the identifications of several emission lines. A direct comparison with
the IUE observations in the range 3045-3300 \AA ~ has revealed the
appearance of many weak and shallow lines, most of them lacking a
convincing identification. The Balmer lines are visible up to H$_{38}$
and are  accompanied by the He II lines of the Pickering series. Also,
all  of the He II emissions of the Pfund series, from 5858 \AA ~to 6408
\AA ~have been detected. We made definite identifications of additional
TiO bands at 4955 \AA ~($\alpha ~R_2 ~ 1-0$), 5167 \AA ~($\alpha~ R_2~
0-0$), 5445 \AA ~($\alpha~ R_2~ 0-1$), 5598 \AA ~($\beta~ R_1~ 0-0$),
5847 \AA ~ ($\gamma '~R_1~ 1-0$),and 6148 \AA ~(${\gamma '} {^S}~R_{21}~
0-0$). The H$\alpha$ line has very wide wings, extending to at least
5000~ km~s$^{-1}$, which are similar to those reported for the planetary
nebula IC 4997 and attributed to Raman scattering by Ly$\beta$  photons.
A selective pumping mechanism via the HeII 237 \AA ~ emission is
proposed to explain the intensity of the high-lying lines of  O IV
mult. 1 and 2.

\end{abstract}

\keywords{08.14.2 novae, cataclysmic variables, 02.12.2 line:
identification - 08.09.2 stars: individual: RR Tel  }
 
\section{Introduction}

The symbiotic nova RR Tel is an extraordinary laboratory for
spectroscopic studies of low-density astrophysical plasmas
on account of the richness of its emission line spectrum that
covers a wide range in ionization and excitation stages.
Since the fundamental study by Thackeray (1977), new optical
observations with higher spectral resolution and S/N ratio have
have gradually improved the quality of the data and
have allowed  the identification of weaker  and
blended spectral features (see Mc Kenna et al 1997, and Crawford
et al. 1999). Also, the presence of a strict correlation 
between the FWHM and the ionization level of the emission lines, 
as pointed out by Thackeray (1977) and confirmed by Penston et al. 
(1983), has provided a simple but powerful tool for the 
identification of spectral lines that has not yet been fully exploited.
  With this in mind, we have taken 
advantage of very recent high resolution VLT--UVES observations of 
RR Tel to revisit the spectral features  of the  nova  and to 
perform an {\it ab initio} identification of its emission features.  
We present here some   highlights of these recent observations, 
deferring  to a next paper a detailed 
description of the spectral identifications and measurements.

\section{Observations and data reduction}

Details on the UVES spectrograph and its performances may be found
in D'Odorico et al (2000) as well as in the UVES User Manual
(D'Odorico \& Kaper, 2000).
The data we used consists of a spectrum obtained on October 16th 1999
with the dichroic \# 1  and the standard setting centered at
3460 \AA ~ in the blue arm and 5800 \AA ~ in the red arm, the exposure time
was of 1200 s for both arms. 
The detector in the blue arm is an EEV CCD, while in the red it is a mosaic
of one EEV (identical to that used in the blue arm) 
and one MIT CCD.
All CCDs are composed of $4096\times 2048$ square pixels of 15 
$\rm\mu m $ side.
The slit width was 0\farcs{6} in the blue
and 0\farcs{8} in the red. 
The data was reduced using the {\tt ECHELLE} context of {\tt MIDAS}
and each CCD was treated independently;
reduction included background subtraction, cosmic ray filtering,
flat fielding, extraction, wavelength calibration and order merging.
Since no arc spectra for wavelength
calibration with this setting are available
for the date of observation,  we  used calibration
spectra acquired on different days. From our previous experience
with UVES (Bonifacio et al 2000), we expect the wavelength scale to
be reproducible  to within  a shift of a few tenths of a pixel;
since we are not interested in accurate radial velocities such a shift
is of no consequence for our analysis.
The resolution, as measured from the Th lines of the calibration lamp
is $\approx 65000$ for the whole blue arm spectrum. For the red arm
spectrum we could not find an arc taken with a slit of 0\farcs{8} and
used one taken with a slit of 0\farcs{66} instead, whose measured
resolution is $\approx 65000$, we therefore expect the resolution
of the red spectra to be slightly lower, i.e. about $60000$, as
predicted by the UVES manual (D'Odorico \& Kaper 2000).
To give an idea of the dynamic range offered by the instrument
we mention that
at the peak of the He II 3203.104 \AA~ line the spectrum detects
7831 $e^-$, while only 88 $e^-$ in the adjacent continuum. An estimate
of the S/N ratio in this range, assuming  Poisson noise, gives S/N
$\sim$ 88 in the He II line and S/N $\sim$ 9 in the continuum.
We also made use of a short exposure spectrum taken 
on October 10th, with the same setting and a slit of 0\farcs{8}
in the blue arm
and 0\farcs{3} in the red arm, the exposure times were 120 s and 60 s
respectively. This spectrum was reduced in the same way as the
long exposure and was used only to check the profiles of
very strong lines.

The measured   wavelength  of the 
sharp, unblended Fe II lines is shifted by $-39.8\pm 1.1$ km~s$^{-1}$
with respect to the laboratory (air) wavelength. After correction 
for the earth motion ($\rm -24.6 ~km~s^{-1}$) the heliocentric  radial
velocity  derived from  the Fe II lines is  $\rm -64.4  ~km~s^{-1}$,
consistent with the 
heliocentric  radial velocity   reported in 
previous works ($\rm -59.7 \pm 1.7$ km~s$^{-1}$, Thackeray (1977)).
Hereafter, we consider the system of the FeII lines as the rest 
frame of RR Tel and the observed wavelengths refer to  a frame that 
is at rest with respect to the Fe II lines.

\section{Results}
\begin{table}
\caption{A selected list of new or newly identified lines in RR Tel}
\begin{center}
\renewcommand{\tabcolsep}{0.1cm}
\begin{tabular}{rllrr}
\hline
$\lambda_{obs}$\phantom{000}\ & ident. & $\lambda_{lab}$   & FWHM
& FWHM \\
  & & & (\AA)
& (km~s$^{-1}$) \\
\hline
 3067.18 &   OVI       &    3066.75    &   0.88  &  86.10 \\
 3138.70 &   [NiVII]   &    3138.3     &   0.65  &  62.10 \\
 3159.14 &   unid.     &               &   0.77  &  73.10 \\
 3221.14 &   [NiVII]   &    3221.5     &   0.53  &  49.40 \\
 3403.51 &   OIV (2)   &    3403.52    &   0.44  &  38.80 \\
 3409.65 &   OIV (3)   &    3409.66    &   0.44  &  38.80 \\
 3411.70 &   OIV (2)   &    3411.69    &   0.44  &  38.70 \\
 3413.61 &   OIV (2)   &    3413.64    &   0.64  &  56.20 \\
 3425.93 &   NeV 1F    &    3425.5     &   0.98  &  85.80 \\
 3428.66 &   OIII      &    3428.63    &   0.36  &  31.50 \\
 3433.69 &   unid.     &    3433.00    &   0.88  &  76.90 \\
 3487.12 &   MgVI      &    3486.7     &   0.87  &  74.80 \\
 3488.90 &   MgVI      &    3488.72    &   0.82  &  70.10 \\
 3502.21 &   MgVI      &    3501.97    &   1.05  &  90.00 \\
 3586.58 &   FeVII     &    3586.32    &   0.72  &  60.20 \\
         & + FeVI      &    3587.66    &         &        \\
 3634.24 &   HeI       &    3634.23    &   0.33  &  27.20 \\
 3725.30 &   CaVI 1F   &    3725.4     &   0.76  &  61.20 \\
         & + OIV (6)   &    3725.81    &         &        \\
 3759.22 &   FeVII     &    3758.92    &   0.84  &  67.10 \\
 4930.55 &   OV (25)   &    4930.27    &   0.89  &  54.10 \\
         & + OIII      &    4930.87    &         &        \\
 5290.54 &   OVI (16)  &    5290.60    &   1.33  &  75.40 \\
         & + OIV (11)  &    5290.1                        \\
 5424.52 &   FeVI 1F   &    5424.22    &   1.05  &  58.10 \\
 5460.68 &   [CaVI]    &    5460.69    &   0.87  &  47.80 \\
 5495.05 &   ArIV] ?   &    5494.39    &   1.59  &  86.80 \\
         & + [CoVIII]  &    5494.8     &         &        \\
         & + FeII 17F  &    5495.82    &         &        \\
 5586.24 &   CaVI      &    5586.26    &   0.90  &  48.30 \\
         & + OIV       &    5585.6     &         &        \\
 5618.60 &   CaVII     &    5618.75    &   1.33  &  71.00 \\
 5631.42 &   CaVI      &    5631.74    &   1.09  &  58.10 \\
         & + FeVI      &    5631.07    &   0.00  &   0.00 \\
 5677.22 &   FeVI 1F   &    5676.95    &   1.12  &  59.20 \\
         & + NII (3)   &    5676.00    &         &        \\
 6086.96 &   FeVII     &    6087.00    &   1.49  &  73.40 \\
         & + CaV 1F    &    6086.37    &         &        \\
 6228.43 &   KVI       &    6228.6     &   1.18  &  56.80 \\
 6500.31 &   OV        &    6500.24    &   0.98  &  45.20 \\
\hline
\end{tabular}

\end{center}
\end{table}

\subsection{The region of overlap with  IUE , 3045-3345 \AA ~}

\begin{figure}
\psfig{figure=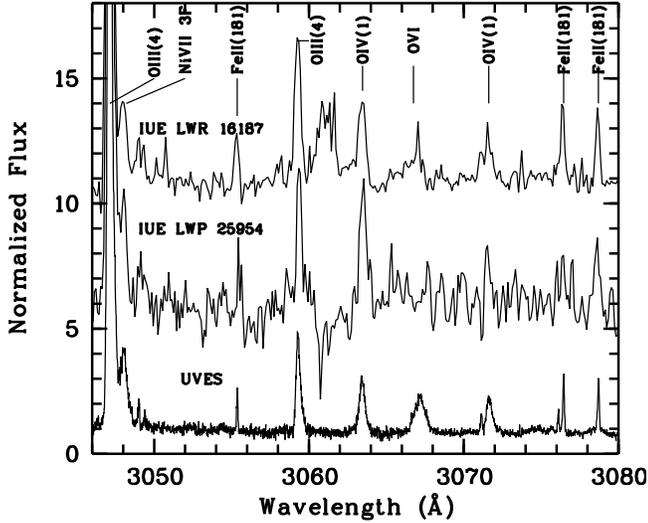,clip=t}
\caption{The UVES spectrum compared to two IUE spectra,
LWR 16187, obtained in 1983 with a 31440 s exposure and
LWP 25954, obtained in 1993 with a 5700 s exposure. All
three spectra have been normalized to the median flux
in the interval 3050-3052 \AA ~ and shifted to rest (air)
wavelength. For display purposes
the two IUE spectra have been shifted vertically by
5 and 10 units.}
\end{figure}

\begin{figure}[t]
\psfig{figure=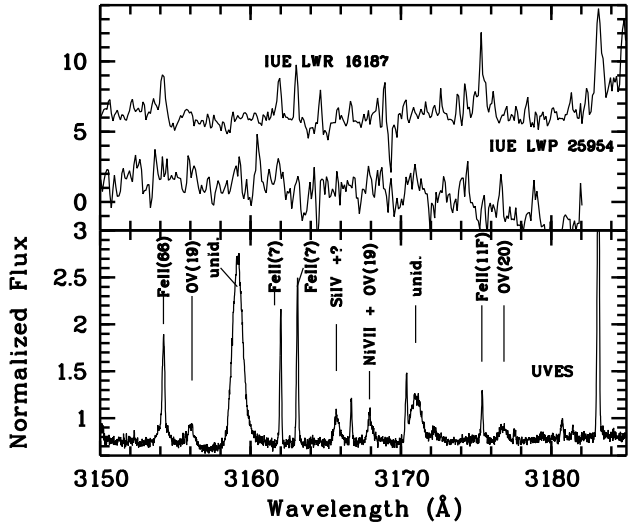,clip=t}
\caption{Comparison of the UVES and IUE spectra in 
the range 3150 - 3185 \AA. The spectra have been normalized and 
shifted as in Fig. 1, but the UVES spectrum is shown on a 
different scale to allow to appreciate the weak features. The tracing
of LWR 16187 has been shifted vertically by 5 units for display
purposes.} \end{figure}

The UVES capability of reaching the region down to  3045 \AA ~
has permitted a direct comparison with the high resolution IUE
observations.
Since the IUE sensitivity declines strongly above 3000 \AA ~, we have
chosen for comparison  the two IUE spectra with the longest
exposure times, [ LWR16187 ($t_{exp}= 31440$ s, June 18,1983 )
and LWP25954 ($t_{exp}=5700$ s, July 20,1993)]. Fig. 1 is a comparison
between UVES and IUE data ( obtained  from the INES archive system
(see Cassatella et al.(2000) and Wamsteker et al.(2000)) in the range
3045--3080 \AA ~ which includes some  identifications.  We point out the
presence in the UVES spectrum of the  wide line (FWHM 86.1 ~km~s$^{-1}$)
at 3066.77 \AA, (identified as OVI 66.75 ) that was marginally present
in LWR16187 and absent in LWP25954, and the lack in the UVES spectrum of
the wide feature near 3061 \AA ~   in LWR16187, which is probably due
to noise.   Fig. 2 is a similar plot for the range 3150--3180 \AA 
~  but the UVES data are on a separate intensity scale to fully 
emphasize their remarkable S/N ratio.    
Several new and wide features are evident  at  3154.19 \AA ~ 
(unidentified, blended with the sharp line of FeII (66) 3154.20
 \AA), 3155.99 \AA ~ (OV(19) 3156.11) , 3159.12 \AA ~ (a strong 
line with FWHM =73.1 ~km~s$^{-1}$ that has defied a convincing
identification) 3165.71 \AA ~  ( SiIV 3165.7 + MnIV // 3165.57 + 
NiVII 3165.4), 3167.91 \AA ~  (OV (19) 3168.10 + NiVII 3168.0), 
3170.97 \AA ~ (unid.),  3176.85 \AA ~  (OV (20) 3176.87).  The two 
sharp lines which are present also in IUE are lines of FeII mult. 
7.  Figure 2 is also a representative illustration of the variety 
of emission profiles that are present in the spectrum of RR Tel: 
we  point out the appearance (or maybe the new detection thanks to  
UVES) of  many weak and shallow lines whose central intensity is 
of the order of  50\% -- 100\% of the continuum. These wide and 
shallow features are quite common in the UV part of the spectrum 
up to  $\lambda 3500$ \AA. Some of these features have been 
identified  as lines of high ionization species  such as OIV-VI 
and  NiIV-VII, but many of them lack proper identification. We 
recall that  Penston et al (1983) listed  20 lines in all in the 
range 3047--3345 \AA ~, including five lines of OIII produced in
the Bowen mechanism. In the UVES spectrum we have detected in the 
same range about 90 emission lines including a dozen of OIII Bowen 
lines. All lines and  identifications of Penston et al (1983) are 
confirmed with two exceptions : the 3063.30 NII (2F) feature is 
missing and TiII 3078.66 \AA ~   is instead  FeII (181) 3078.69 
(see also Fig. 1);  no TiII line has been positively   detected in 
our spectra.   Among other new features in the 
IUE range  we mention also the lines at 3131.21 \AA ~  ( NiVI 
3131.4), 3138.66 \AA ~  ( NiVII 3138.3), on the wings of the 
strong OIII(12) 3132.86 \AA ~  line  , and 3221.12 \AA ~ ( NiVII 
3221.5 ).  In the context of the application of new laboratory 
analysis of the spectrum of highly ionized nickel to the 
identification of unclassified lines in the spectra of $\eta$  Car
and RR Tel, the NiVII 3221.5 \AA ~  line was mentioned by Raassen and
Hansen (ApJ 243,217, 1981) as a forbidden line falling in a region 
difficult to observe.

\subsection{The 3350--3870 \AA ~  region}

\begin{figure*}[t]
\psfig{figure=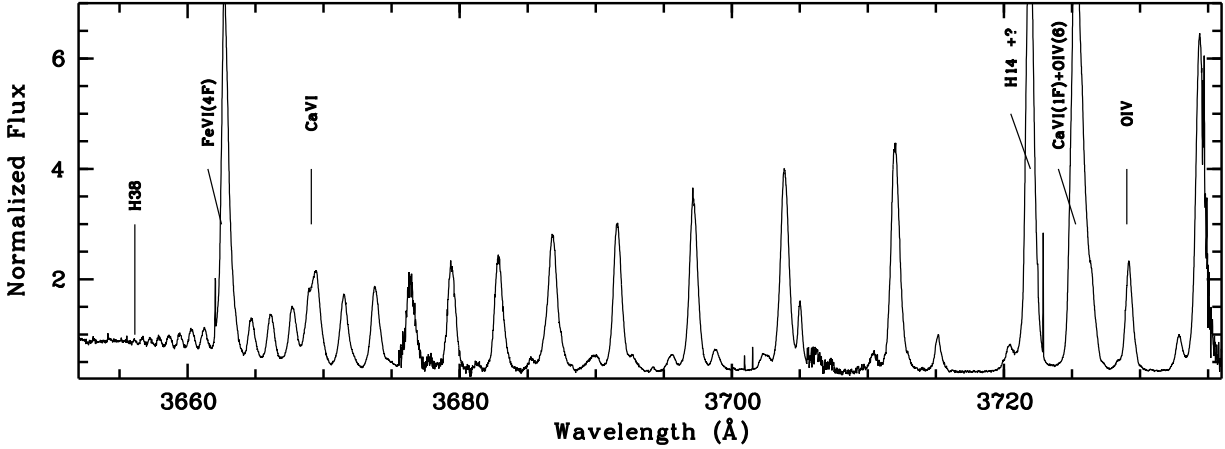,clip=t}
\caption{The Balmer discontinuity and the last members
of the Balmer series. The spectrum has been normalized
as in figures 1 and 2. The spectral regions with larger noise
correspond to the intervals where overlapping orders have been merged.}
\end{figure*}

An obvious comparison  here is with the recent study by Crawford 
et al.(1999) on AAT--UCLES data of slightly lower  resolution than
the VLT-UVES data.   We are aware that the identification of emission 
lines is a somewhat arbitrary procedure  but, on the basis of the 
FWHM criterion and of common astrophysical sense,  we disagree 
with Crawford et al. on the  identifications of low ionization
species such as TiI, VII, MnI, MnII etc. 
We have found  alternative identifications generally 
associated with higher ionization species.  We postpone  a 
detailed  discussion on line identifications  to the follow-up  
paper and  we just  give here, in  Table 1,  a short
list of strong lines either not reported or incorrectly identified 
by Crawford et al., together with our proposed  identifications.  
Thanks to the very high spectral 
resolution of the UVES spectrograph we have been able to detect 
emission lines of  the Balmer series  down to H38 at 3656.13 \AA ~  
(see Fig 3). Also, starting from approximately H18, where the  
separation between successive lines becomes large enough, it is 
clearly seen that ALL  Balmer emissions are accompanied by a 
shortward displaced, much  weaker emission that we associate with 
HeII lines of the Pickering (4-n) series. This is confirmed by the 
presence of the other  HeII lines of the same series that fall in  
between the Balmer lines. 
Crawford et al discarded a possible  contribution of [SIII] 
3721.70 \AA ~  to the H14 Balmer line at 3721.94 \AA, as in Mc 
Kenna et al (1997), on the basis of the FWHM of the feature, which  
agreed well with that of other H lines. As a matter of fact, the 
feature we observe at 3721.89 \AA ~  (see also
Fig. 3) has the same FWHM (55 ~km~s$^{-1}$) as the other H lines, but
clearly shows an intensity  excess with respect to the adjacent H
lines that requires an additional contribution at a wavelength 
necessarily very close to that of the H14 line.
   The presence of the two OII 1F  lines at 3726.03 \AA ~ and 
3728.81 \AA ~  is  questionable: we recall that Contini and 
Formiggini (1999) from the presence of [OII]3727 \AA ~  inferred 
an electron density in the emitting gas not much higher  than 
$3\times 10^3\rm cm^{-3}$, a disturbingly low value that is in 
contrast  with the densities derived by  other diagnostic methods. 
The observed features (see Fig. 3) fall at 3725.27 \AA ~  and at 
3729.17 \AA ~ with FWHM of  63 ~km~s$^{-1}$ and 42.6 ~km~s$^{-1}$ ,
not compatible  with  OII since  
the OIII lines have FWHM close to 30 ~km~s$^{-1}$. We identify the
former line as [CaVI] 1F 3725.4 \AA ~  + OIV (6)3725.81 \AA ~ (already
in  Thackeray (1977)),  while the latter line is OIV 3729.03 \AA  ;
its FWHM (42.6 ~km~s$^{-1}$) is in  good  agreement with the FWHM of
other OIV lines  (~ 41 ~km~s$^{-1}$ ). Finally,  the 3710 \AA ~
SIII line reported in Table 5 of  Contini and Formiggini, is, more 
likely, HeII 3710.44 \AA ~  (4-30), the shortward $\lambda$ 
companion to the H15 emission line.

\subsection{The  4780 - 6820 A Region}

\begin{figure}[t]
\psfig{figure=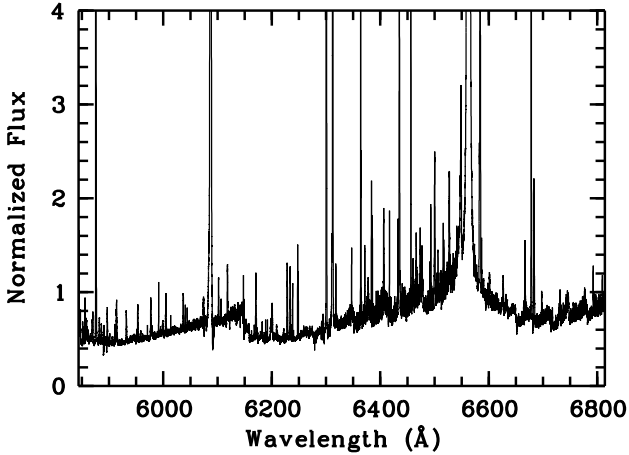,clip=t}
\caption{The whole range covered by the upper CCD of the red arm.
The flux has been normalized as in the previous figures and the 
wavelength shifted to rest. The wide wings of H$\alpha$ are likely
due to Rayleigh--Raman scattering.  }
\end{figure}

Some strong emission lines 
either not listed or incorrectly identified  by Crawford et al.(1999)
are reported in Table 1.
The same considerations given in the previous Section  apply 
here to rule out  identifications based on  low ionization 
species such as Ca I, TiI, MnI, Mn II, VII, etc.
 We have identified ALL of the HeII emission of the Pfund 
(5-n) series, from  5858.88 \AA ~  (5-30) to 6408.15 \AA ~ (5-15). Some 
of the higher members of  this series have been identified as 
lines of low ionization species  in previous works. 
 
If one looks only to the details of the individual emission 
lines there is the  risk of missing  the forest for the trees; 
instead,  from a general view of the entire spectrum at  a proper 
scale, two interesting  features come out:

1.  The H$\alpha$ line has very wide wings that extend from about
6300 \AA ~ to about 6700 \AA. (see  Fig 4). While a crowding of emission
lines can partially contribute to the violet wing, the red wing
is relatively free  of  contamination  and indicates  a velocity  
on the order of at least 5000 ~km~s$^{-1}$ . Similar broad wings have
been reported for the planetary nebula IC 4997 by Lee and Hyung (2000)
who attributed them  to Rayleigh-Raman scattering  by 
which Ly$\beta$ photons with a velocity width  of a few $10^2$
km~s$^{-1}$ are converted to optical photons and fill the H$\alpha$
broad wing region. We recall that Schmid (1989) and Schmid and Schild
(1990) interpreted  two broad features  at 6825 and 7082 \AA ~ in RR Tel
and other symbiotic stars as the Raman scattered UV OVI 1032 \AA ~
and 1038 \AA ~  resonance lines. Van Groningen (1993) detected  
Raman scatttered HeII(2-8) and (2-10) at 4333 \AA ~ and 4851 \AA;
this latter line is also present in our spectrum as a wide and 
shallow feature.

2. Webster  (1974) reported some evidence
of the presence of TiO absorption bands in the red spectrum of RR
Tel. This  might testify the contribution to the spectrum by the Mira
variable. Mc Kenna et al (1997) found no sign of
these TiO absorption bands, but Crawford et al (1999) made definite
identifications of the TiO absorption bands near 
6651 \AA ($\gamma ~R_3~ 1-0$),7052 \AA ($\gamma ~R_3 ~ 0-0$9, 
7666 \AA ($\gamma ~ R_3 ~1-2$) and
8206 \AA ($\delta ~R_1 ~ 1-0$).  
Figures 4 and 5  clearly show the presence of additional  TiO
bands at 4955 \AA ($\alpha ~R_2 ~ 1-0$),  
5167 \AA ($\alpha~ R_2~  0-0$), 5445 \AA ($\alpha~ R_2~ 0-1$), 
5598 \AA ($\beta~ R_1~ 0-0$),
5847 \AA ($\gamma '~R_1~ 1-0$),and 6148 \AA (${\gamma '} {^S}~R_{21}~
0-0$). For the band classification we used the list of bandheads
provided in the Appendix of Valenti et al (1998).
There is also evidence for weaker absorption
bands  (including that at 6651 \AA) in the region longward
of H$\alpha$.

\begin{figure*}[t]
\psfig{figure=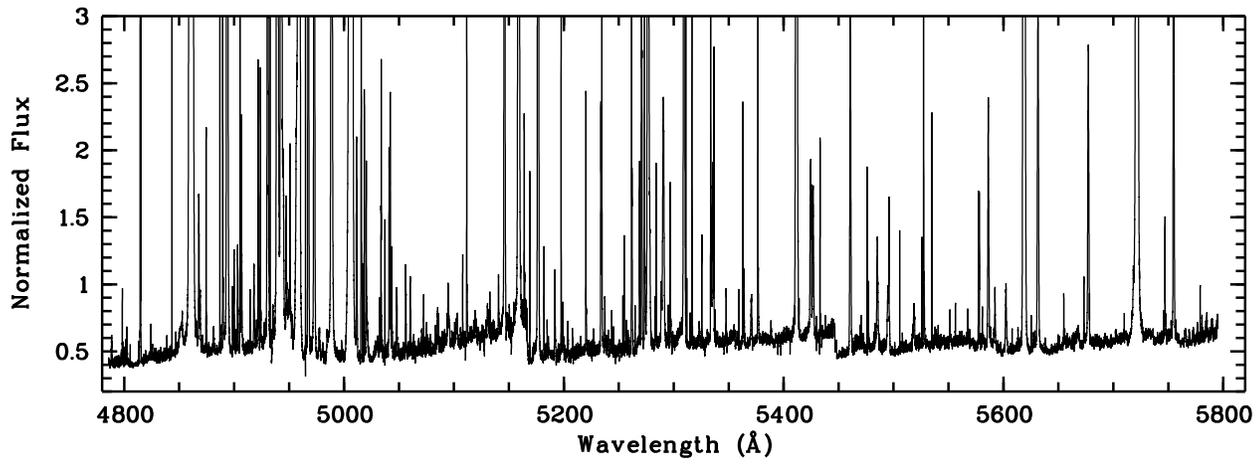,clip=t}
\caption{The whole range covered by the lower CCD of the red arm.
The flux has been normalized as in the previous figures and the
wavelength shifted to rest. The jumps due to TiO are
readily visible.}
\end{figure*}

\subsection{A selective  pumping of the optical OIV lines ?}

 The optical OIV  lines come from  levels with energies
above  48.3 eV and therefore, in contrast with the UV lines
near 1400 \AA,  collisional excitation is unlikely.
 The lines of OIV mult. 1, 2, 3, and 6  are present as a
 clearly evident  features in the UVES spectrum of RR Tel, while
the lines of  mult.  4,5,7, 8, which have comparable excitation and
intensities  are nearly absent. Thus, a mechanism of selective
excitation seems at work.
 The lines of mult. 2 near 3400 \AA ~(which decays into the lines of
mult. 1 near 3070 \AA ~) have the upper term ( at 52.02 eV.) in common
with the two lines at 238.57 and 238.36 \AA ~  of mult. UV5.
This  wavelength is rather close to that of a HeII Lyman
transition at 237.33 \AA, (arising from a level at  52.24 eV) and
we suggest that the HeII  237 \AA ~  line   pumps  the upper term of
OIV mult. UV5 which in turn decays into mult. opt. 2 and opt. 1 and
other lines.
  Apparently a similar mechanism should work for the upper level of
mult. 6 at 61.40 eV (which in turn decays into mult. opt  3) but we have
found no candidates for the pumping line.

\begin{acknowledgements}

We are grateful 
to S. D'Odorico, H. Dekker and the whole UVES team for
conceiving and building such a terrific instrument as UVES. We are
indebted with M. Tarenghi for suggesting RR Tel as a target for the
commissioning observations.  We made large use of the excellent Atomic
Line List v.2.04 maintained by Peter van Hoof at the WEB site
http://www.pa.uky.edu/$\sim$peter/atomic . \end{acknowledgements}


\begin{thebibliography}{}

\bibitem{} Bonifacio P., Hill V., Molaro P., Pasquini P., Di Marcantonio P.,
Santin P., 2000, A\&A 359,663
\bibitem{} Cassatella A., Altamore A., Gonzalez-Riestra R., Schartel N.,
  Wamsteker W.,2000, A\&AS,141,331
\bibitem{} Contini M. and Formiggini L.,1999 Ap.J 517,925
\bibitem{} Crawford F.L., McKenna F,C., Keeenan F.P., Aller L.H., Feibelman
  W.A., Ryan S.G.,1999, A\&AS,139,135
\bibitem{} D'Odorico S., Kaper L. "UV-Visual Echelle Spectrograph User
Manual"  2000, Doc. No. VLT-MAN-ESO-13200-1825,
http://www.eso.org/instruments/uves/userman/
\bibitem{} D'Odorico S., Cristiani S., Dekker H., Hill V., Kaufer A., Kim
T.,  Primas F.,2000, SPIE Proceedings 4005, in press
\bibitem{} Lee H., Hyung S.,2000, ApJ530,L49
\bibitem{} McKenna F.C., Keenan F.P., Hambly N.C.,
Allende Prieto C., Rolleston W.R.J., 1997,ApJS,109,225
\bibitem{} Penston M.V., Benvenuti P., Cassatella A., Heck A.,
 Selvelli P.L., Macchetto F., Ponz D., Jordan C.,  Cramer N., Rufener
F., Manfroid J.,1983, MNRAS 202,833
\bibitem{} Raassen A.J.J. and Hansen, J.E.,1981, ApJ243,217
\bibitem{} Schmid H.M.,1989, A\&A,211,L31
\bibitem{} Schmid H.M., Schild,H.,1990, A\&A,236,L13
\bibitem{} Thackeray A.D.,1977, Mem.RAS,83,1
\bibitem{} Van Groningen E.,1993, MNRAS,264,975
\bibitem{} Valenti J.A., Piskunov N., Johns-Krull C.M., 1998, ApJ 498, 851
\bibitem{} Wamsteker W., Skillen I., Ponz J.D., de la Fuente A., Barylak
M.,and Yurrita I.,2000, Ap\&SS,in press
\bibitem{} Webster B.L.,1974, IAU Symp.59,123




\end{thebibliography}
\end{document}